\documentclass{cimento}
\title{Solar Convection and Magneto-Convection Simulations}

\author{
Robert F. Stein\from{ins:msu},
David Bercik\from{ins:msu},
Aake Nordlund\from{ins:tac}
}

\instlist{\inst{ins:msu} Michigan State University, East Lansing, MI, USA
\inst{ins:tac} Niels Bohr Institute and Theoretical Astrophysics Center, Copenhagen, DK }

\input epsf
\input psfig

\begin{document}

\maketitle


\begin{abstract}
Magneto-convection simulations with two scenarios have been performed:
In one, horizontal magnetic field is advected into the computational
domain by fluid entering at the bottom.  In the other, an initially
uniform vertical magnetic field is imposed on a snapshot of
non-magnetic convection and allowed to evolve.  In both cases, the
field is swept into the intergranular lanes and the boundaries of the
underlying mesogranules.  The largest field concentrations at the
surface reach pressure balance with the surrounding gas.  They
suppress both horizontal and vertical flows, which reduces the heat
transport.  They cool, become evacuated and their optical depth unity
surface is depressed by several hundred kilometers.
Micropores form, typically where a small granule disappears and
surrounding flux tubes squeeze into its previous location.
\end{abstract}

We have been making realistic simulations of meso-granule scale
convection near the solar surface.  We begin with a summary of three
basic properties of stratified convection.  Next, we discuss our
magneto-convection results.  Finally, we present results on micropore
structure and formation.

\section{Convection: Basic Properties}

First, convection is driven from a very thin surface thermal boundary layer
where radiation cools the fluid and
produces the low entropy fluid which is pulled down by gravity
and forms the cores of the cool downdrafts
where most of the buoyancy work that drives the convection
occurs.

Second, convective flow is controlled by mass conservation.  Fluid moving
toward the surface in a stratified atmosphere, where density decreases
upward, has to turn over and head back down in a distance the order of a
scale height.  Therefore upflows are diverging and fairly laminar, while
downflows are converging and turbulent.  As a result, only a small fraction
of fluid at depth reaches the surface and there is little recycling
of downflows into upflows.  The horizontal scale of the upflows
decreases toward the surface as the scale height decreases in
proportion to the temperature.
\begin{figure}[htb]
 \centerline{\psfig{figure=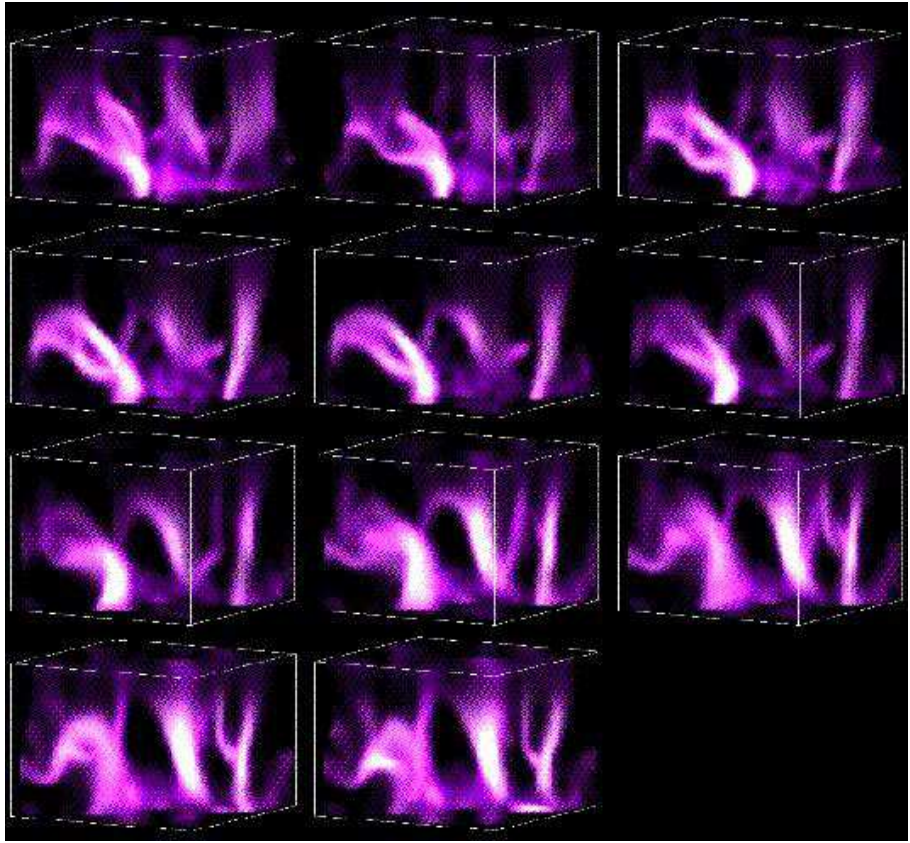,width=12cm}}
 \caption[]{\label{loops-t}
 {\small Magnetic field loops rising through the solar surface and opening
 up.  Time sequence is across then down.  The upper boundary
 condition is a potential field.  The domain is 6 $\times$ 6 Mm
 $\times$ 3 Mm deep.}
 }
\end{figure}

Third, near the surface of the Sun hydrogen is partially ionized and
most of the energy is transported as ionization
energy which gets dumped at the surface when hydrogen becomes
neutral.  The downward kinetic energy flux is of order 10-15\% of the
enthalpy flux.  The radiative flux is negligible below the surface.
(In simulations with a large radiative flux in
the interior. the temperature fluctuations are reduced and therefore
larger velocities are required to carry a given flux.  In such
calculations there is a large downward kinetic energy flux.)

Detailed results and comparisons with observations are
given in Stein and Nordlund (1998, 2000).
\begin{figure}[htb]
  \hbox{
  \epsfxsize=6.5cm
  \epsffile{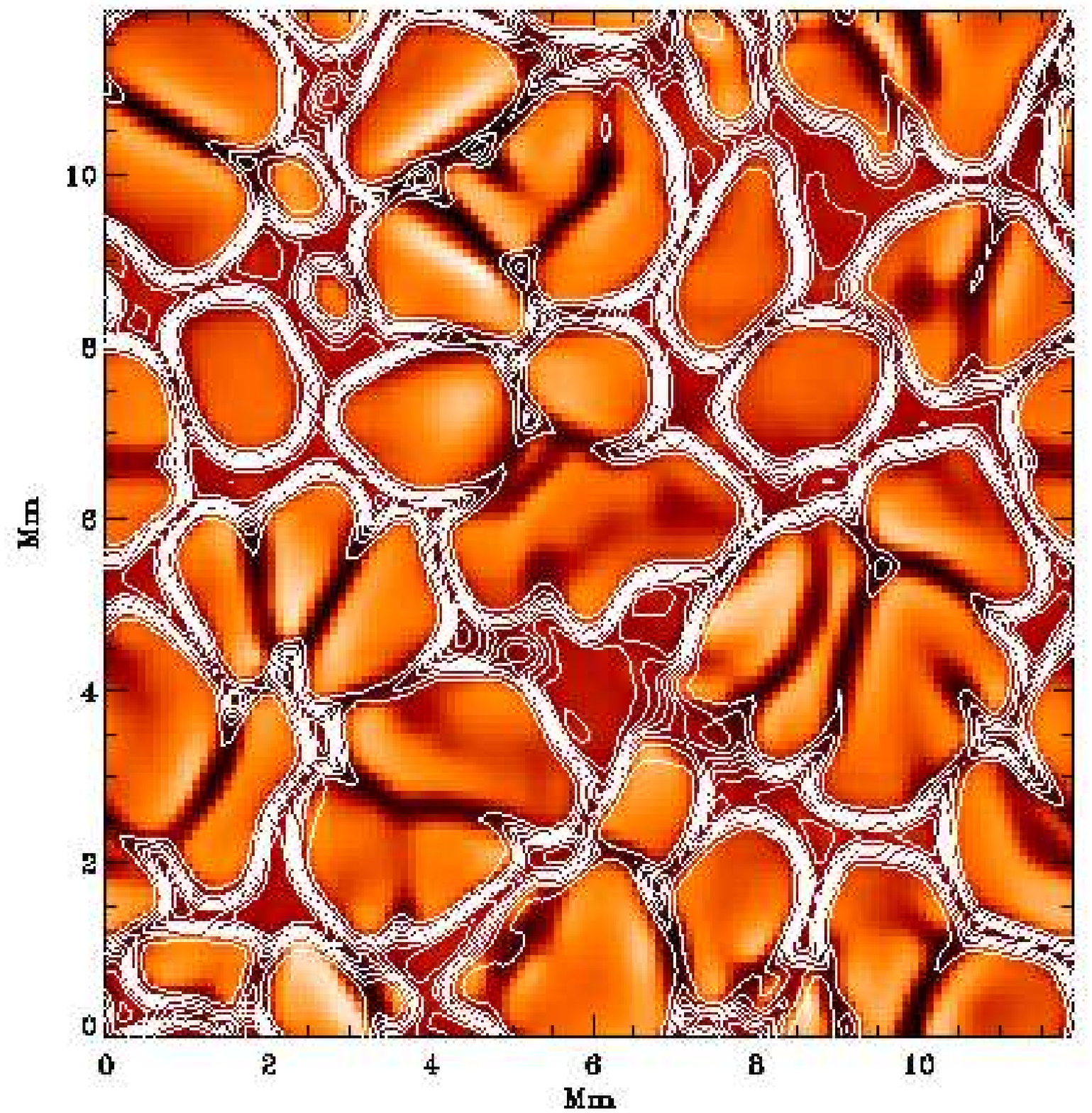}
  \hspace*{0.3cm}
  \vspace*{-0.2cm}
  \epsfxsize=6.5cm
  \epsffile{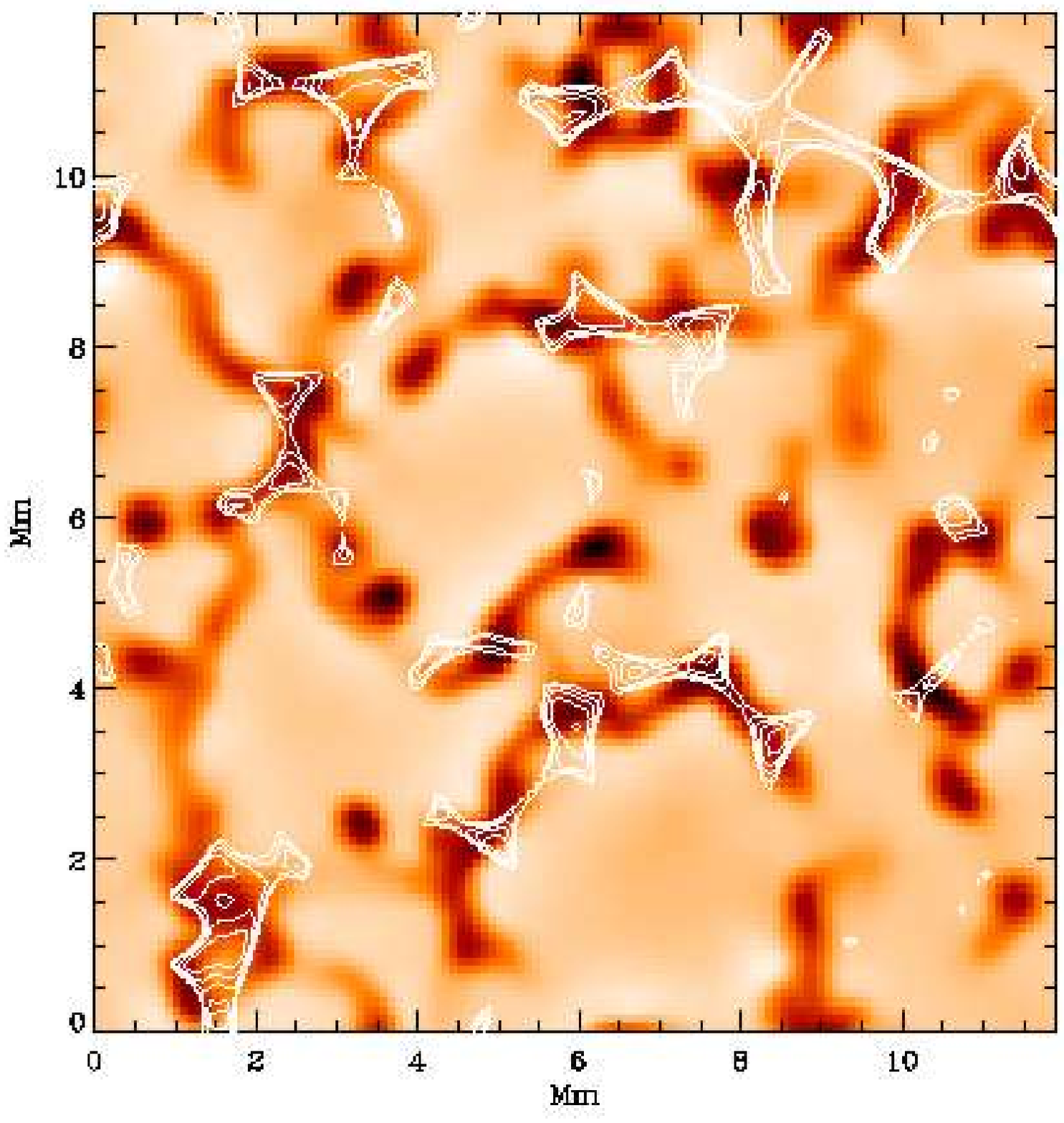}
  \vspace*{0.3cm}
  }
  \hbox{
  \parbox[t]{6.6cm}{
  \caption[]{\label{Vz-B-surf}{\small
  Image of vertical velocity with magnetic field contours.  Magnetic
  field is confined to the intergranular downflow lanes, but does not
  entirely fill them.  The domain is 12 $\times$ 12 Mm $\times$ 3 Mm
  deep.}
  }}
  \hspace{0.2cm}
  \parbox[t]{6.6cm}{
  \caption[]{\label{Vz-B-bot}{\small
  Image of vertical velocity at depth 2.5 Mm with contours
  the surface magnetic field.  The magnetic field collects
  in the mesogranular boundary downflows.}
  }}
  }
\end{figure}
\begin{figure}[htb]
  \centerline{\hbox{
  \epsfxsize=5.0cm
  \epsffile{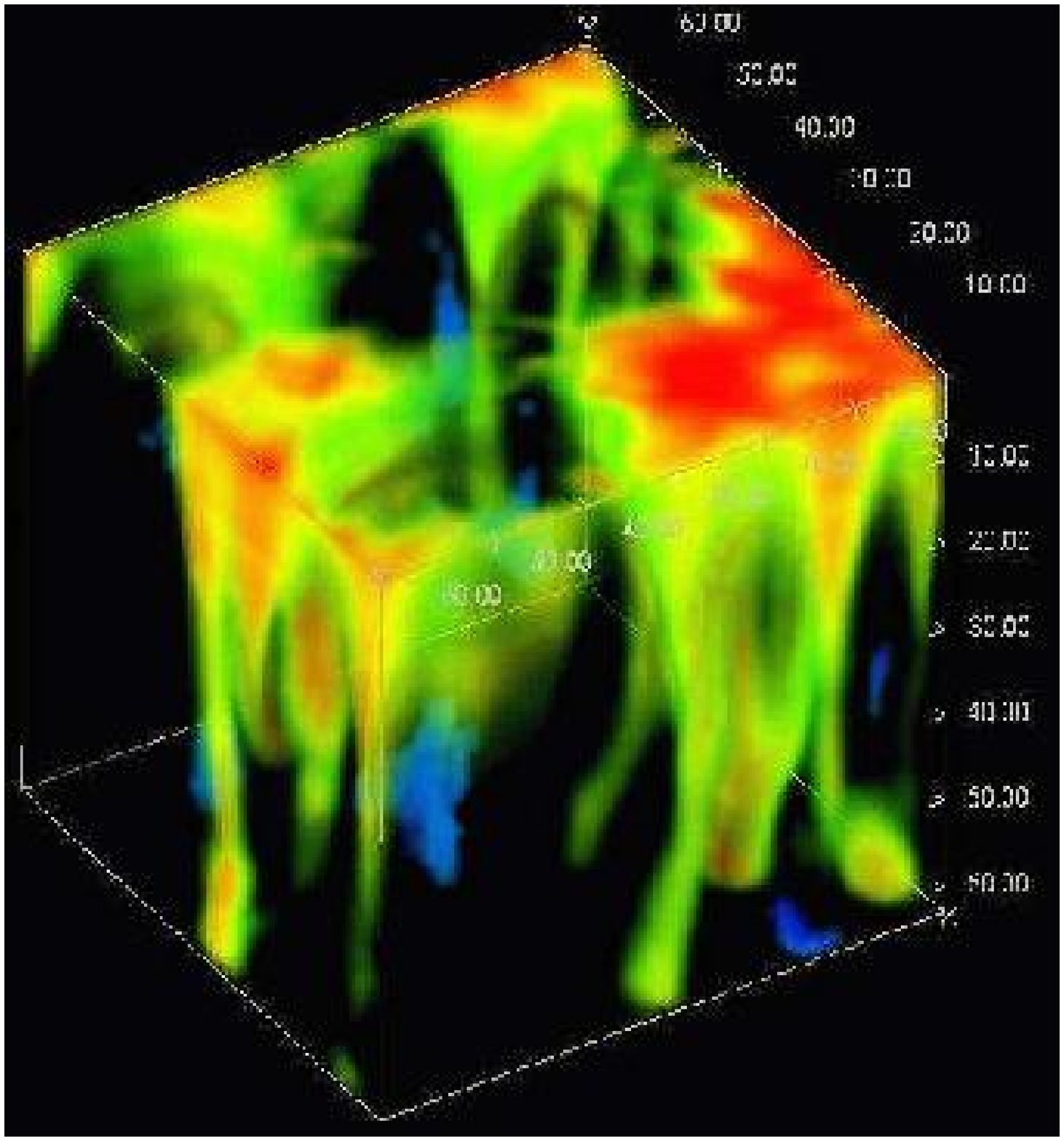}
  \hspace*{0.3cm}
  \vspace*{-0.2cm}
  \epsfxsize=6.5cm
  \epsffile{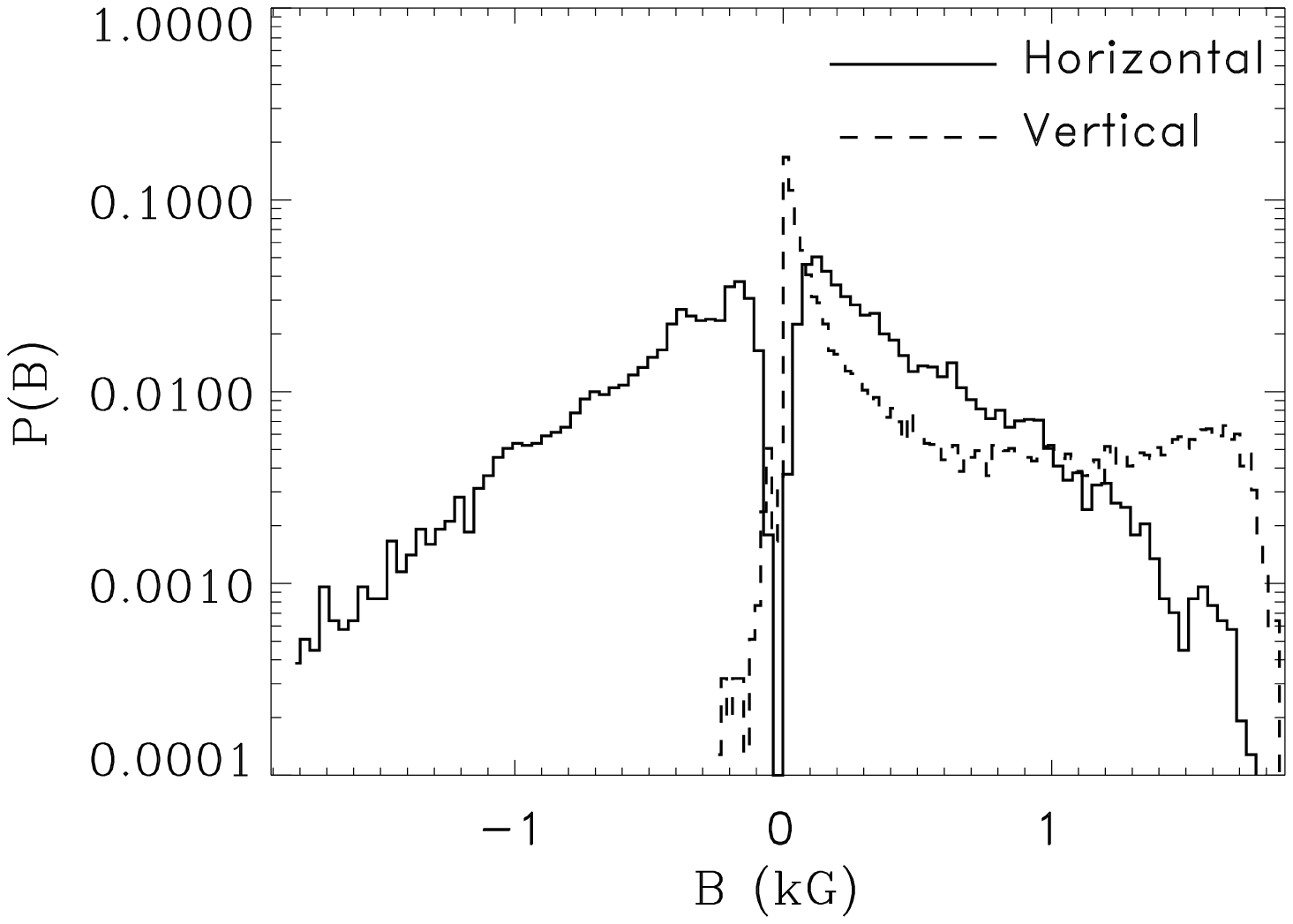}
  \vspace*{0.3cm}
  }}
  \centerline{\hbox{
  \parbox[t]{5.0cm}{
  \caption[]{\label{B3D}{\small
  3D image of magnetic field. Magnetic flux tubes have a
  filamentary structure.}
  }}
  \hspace{0.2cm}
  \parbox[t]{7.0cm}{
  \caption[]{\label{b_hist-v-h}{\small
  The surface magnetic field has an exponential distribution for
  the case of a horizontal field advected in from the bottom, and it
  has a long tail with a bump near the value for pressure
  equilibrium with its surroundings for the case of a uniform vertical
  field.}
  }}
  }}
\end{figure}

\section{Magneto-Convection}

One of magneto-convection simulation looked at the emergence of
magnetic flux through the solar surface.  Horizontal
magnetic field was advected into the computational domain by the fluid
flowing in through the bottom boundary.   The magnetic flux rises to
the surface from a depth of 2.6 Mm in about one hour.  A time
sequence of images of the magnetic intensity is shown in Fig.~\ref{loops-t}
(view across then down).  A flux loop emerges through the surface
and then opens out (snapshots 5-11 center) and another loop
opens up for a while and then closes in on itself and is pulled
back down again below the surface (snapshots 8-11 left).

Another of magneto-convection simulation was started from a snapshot of
the non-magnetic convection, on which we
superimposed a uniform vertical field of 400G and let it
evolve.  (This is high even for a plage region, but we wanted enough
flux to form pores.)  As one would expect, the magnetic field gets swept
into the intergranular lanes, and
tends to concentrate above the boundary of the underlying mesogranule
cells near the bottom of the domain (Figures~\ref{Vz-B-surf} and
\ref{Vz-B-bot}).
Even with this large an average field, the magnetic flux does not fill
all the intergranular lanes.  The magnetic flux tubes suppress both the
vertical and horizontal fluid flow.  As a result, the convective heat
transport is suppressed and the flux tubes become cool and dark.  They therefore
become evacuated.  Both because of their lower temperature and lower density,
their optical depth unity surface gets depressed by several hundred
kilometers.
Above the surface the field spreads out because $\beta$ is
smaller than one.
Below the surface, the magnetic field is filamentary.
A strong flux concentration at the surface separates into several
individual flux tubes below the surface (Fig.~\ref{B3D}).

The magnetic field at the surface has an exponential
distribution for the case where horizontal field is advected into the
domain from below (Fig.~\ref{b_hist-v-h}).
For the case of an initial uniform vertical field, there is
a very small exponential component and a large one sided tail with a
peak at 1.7 kG, near the value for pressure equilibrium with its
surroundings.
\begin{figure}[htb]
  \hbox{
  \epsfxsize=6.5cm
  \epsffile{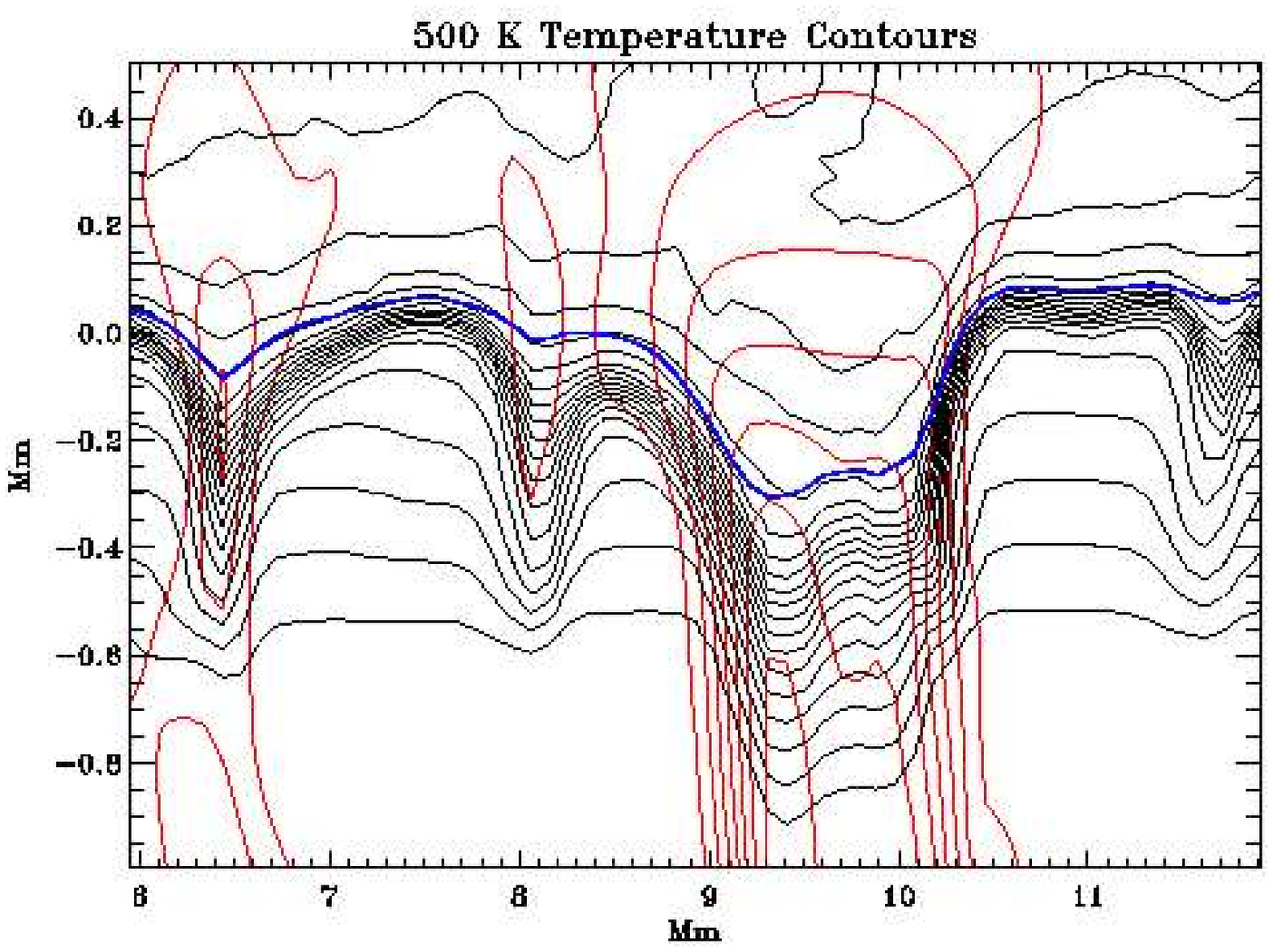}
  \hspace*{0.3cm}
  \vspace*{-0.2cm}
  \epsfxsize=6.5cm
  \epsffile{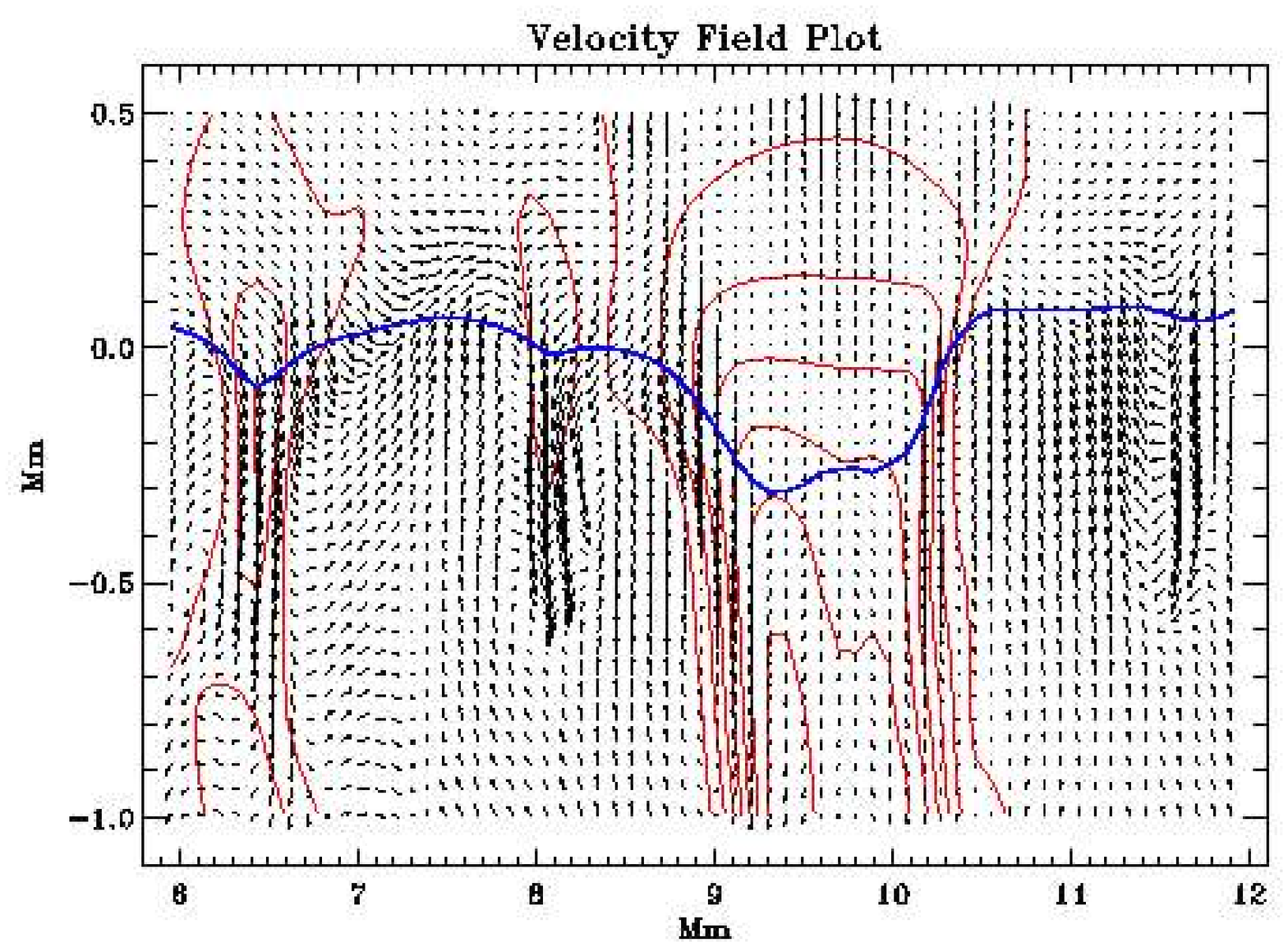}
  \vspace*{0.3cm}
  \hspace*{-14.0cm}
  \parbox[t]{14.0cm}{
  \caption[]{\label{T-B-pore}{\small
 Left: Temperature (dark) and magnetic field (grey) contours in bright point (left)
 and micropore (right).  Temperature is lower in the bright point and
 micropore.  Heavy line is optical depth unity, which is depressed in the
 both the bright point and micropore.  The $\tau = 1$ level cuts the
 bright point at a higher temperature than the surroundings and cuts the
 micropore at a lower temperature than the surroundings.
 Right: Velocity vectors and magnetic field contours in bright point (left)
 and micropore (right).  The flow is suppressed in the micropore, but
 not in the bright point.}
 }}
 }
\end{figure}

\section{Micropores}

\begin{figure}[htb]
 \centerline{\psfig{figure=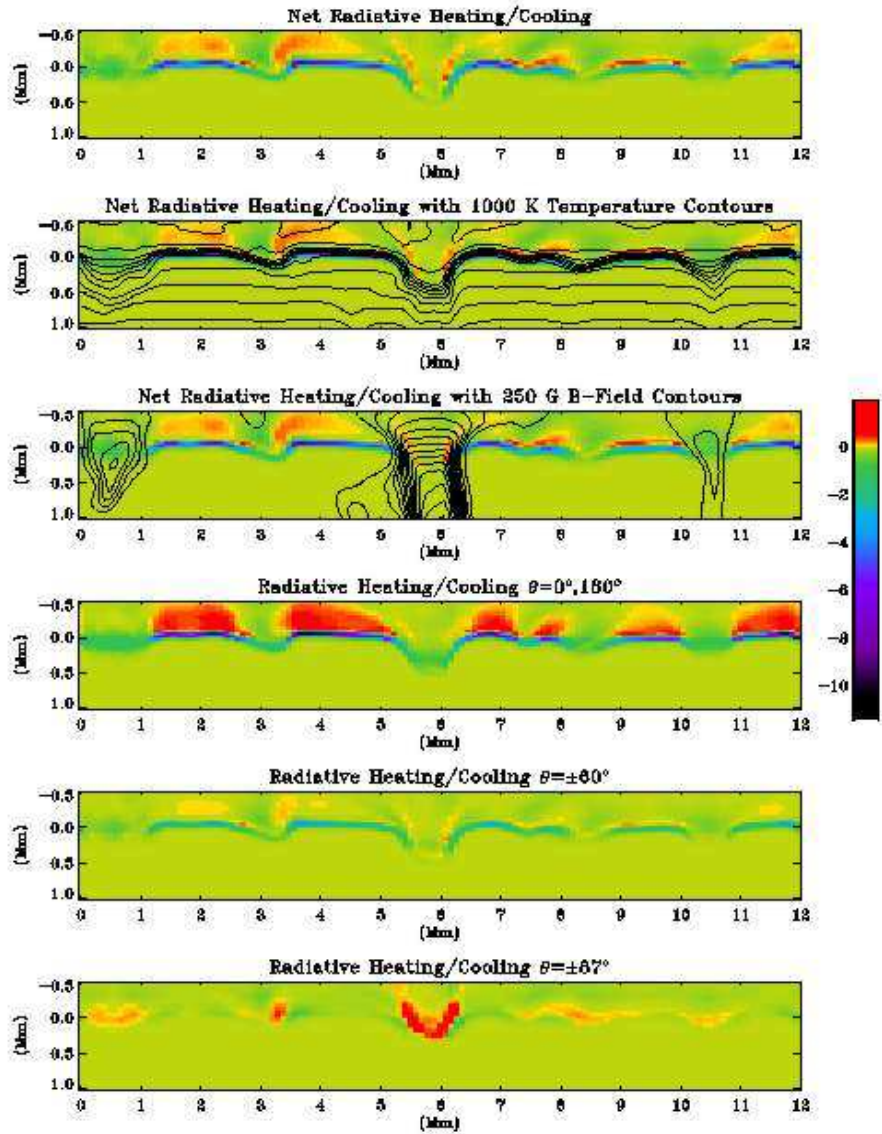,width=12cm}}
 \caption[]{\label{Qrad}
 \small{Radiative heating (red) and cooling (blue and purple).
 Micropores cool vertically and are heated from their hot sidewalls.}
 }
\end{figure}

Micropores form in the simulation at the vertices of the intergranular
lanes.  They evolve on a granulation time scale, with only a few
lasting several granule turnover times.
Micropores form in strong magnetic field concentrations where both the optical
depth unity surface and the temperature contours get depressed
(Fig.~\ref{T-B-pore}). In a
micropore the optical depth unity surface cuts the temperature contours
at a lower temperature than the surroundings, which is why they look
dark.  The whole structure is depressed because it is cool,
has a smaller scale height, and so becomes partially evacuated.  Where
the field concentration is small, in a downdraft,
the velocity is still strong.  But in the micropores the flow is
confined to its edge (Fig.~\ref{T-B-pore}).  In the center of the
micropore where the field
is strongest the velocity is severely suppressed, and therefore
the convective energy transport is severely suppressed.
The micropores are cooled by radiation in the vertical direction and heated
by radiation from their hot sidewalls (Fig.~\ref{Qrad}).  The vertical cooling
is nearly balanced by the horizontal heating and the micropore structure is
fairly stable.
\begin{figure}[htb]
 \centerline{\psfig{figure=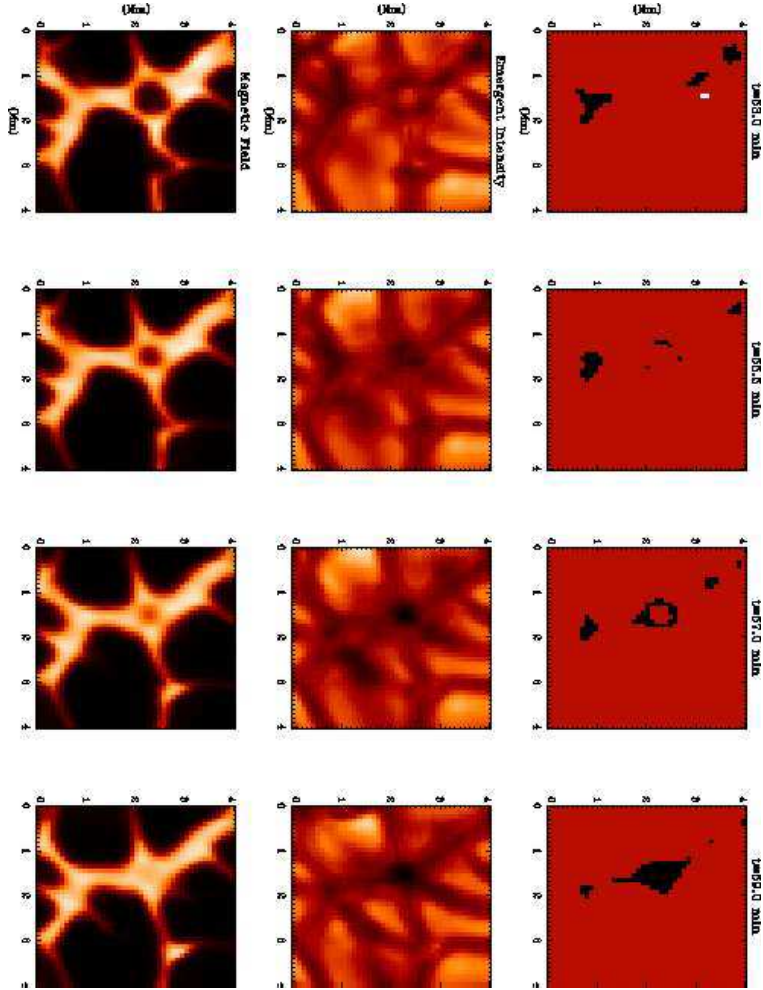,width=10cm}}
 \caption[]{\label{B-I-mask}
 \small{Formation of a micropore.  Magnetic field (left), Intensity (center),
mask showing only the strongest magnetic field (right).
A tiny granule is squeezed out of existence as the
magnetic field initially forms a ring around it and then concentrates
at the intergranular lane vertex when the granule disappears.}
 }
\end{figure}
\begin{figure}[htb]
 \centerline{\epsfxsize=12cm\epsffile{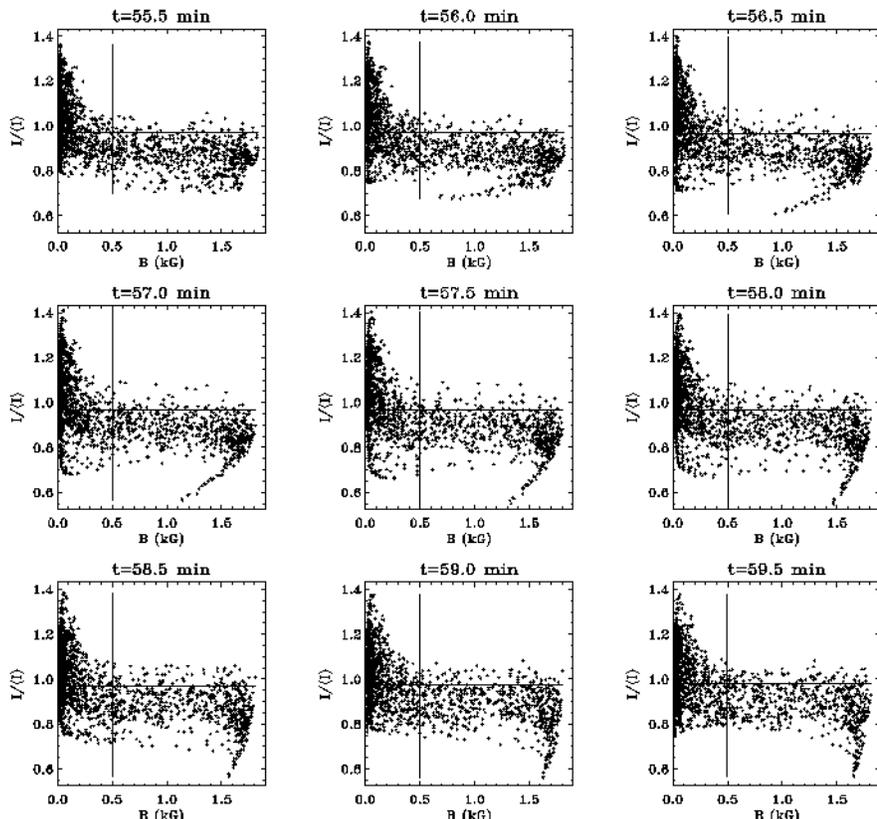}}
 \vspace*{-0.5cm}
 \caption[]{\label{IvsB-time}
 \small{Intensity vs. magnetic field strength during pore formation.  The
dark pore forms first where the field is weak and the field strength
gradually increases over time.}
 }
\vspace*{-0.8cm}
\end{figure}

Micropores tend to form where a tiny granule disappears.
The magnetic field initially consists of
several flux tubes in a ring around the vanishing granule.  When the
granule disappears, the area becomes dark and the field converges
into the area formerly occupied by the granule (Fig.~\ref{B-I-mask}).
As the field moves inward, the intensity
drops even further.  The strong magnetic field eventually occupies the
entire micropore area at the vertex of several intergranular lanes.  Another
way of looking at the same process is via the correlation of the emergent
intensity with the magnetic field
(Fig.~\ref{IvsB-time}).
As the micropore forms the darkest region initially has low field strength.
As time goes on, the magnetic field strength at the darkest points in
the micropore increases, until it reaches the maximum field strength
around 1.6 kG.

\begin{figure}[htb]
  \hbox{
  \epsfxsize=6cm
  \epsffile{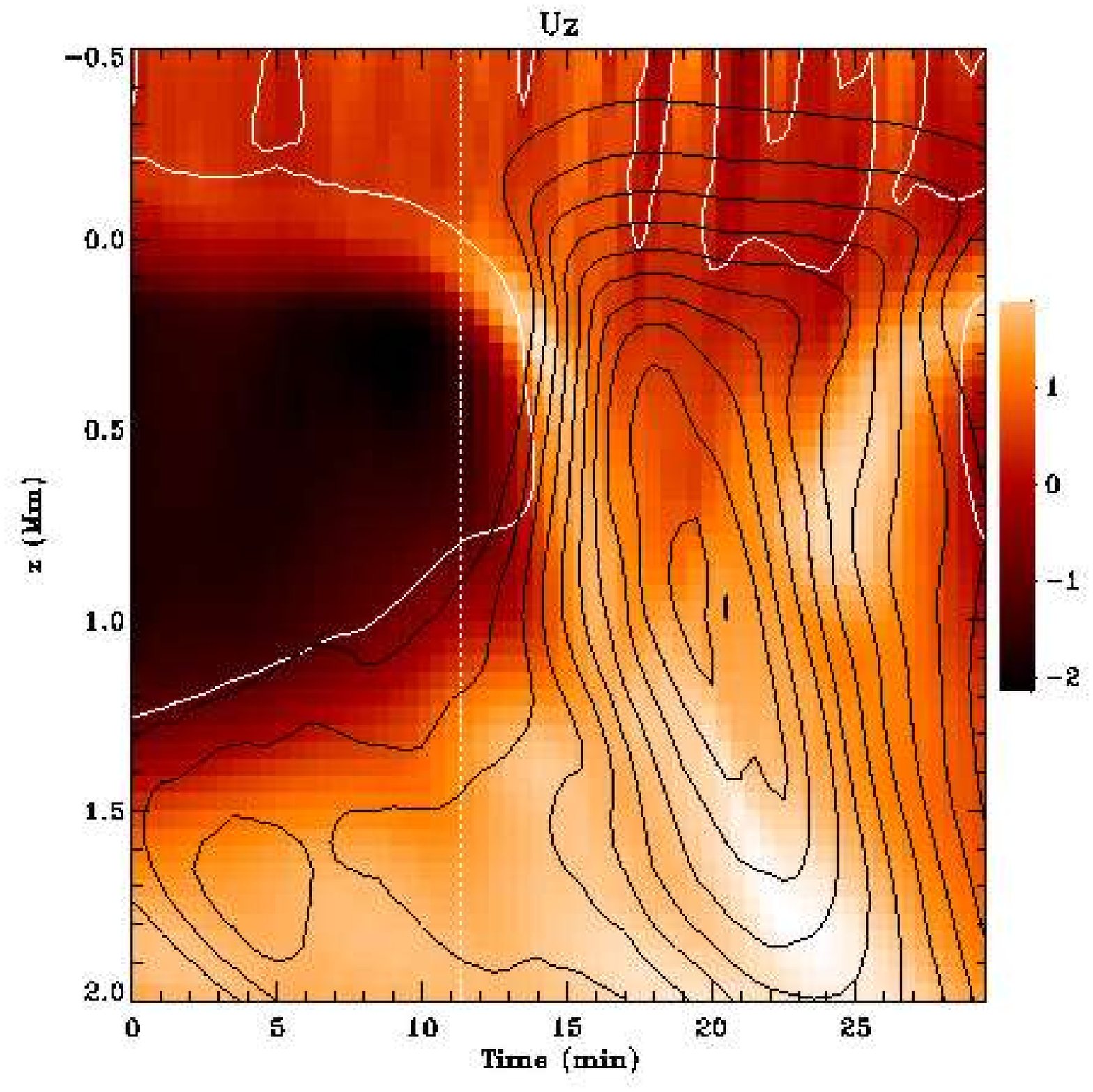}
  \hspace*{0.3cm}
  \vspace*{-0.2cm}
  \epsfxsize=6cm
  \epsffile{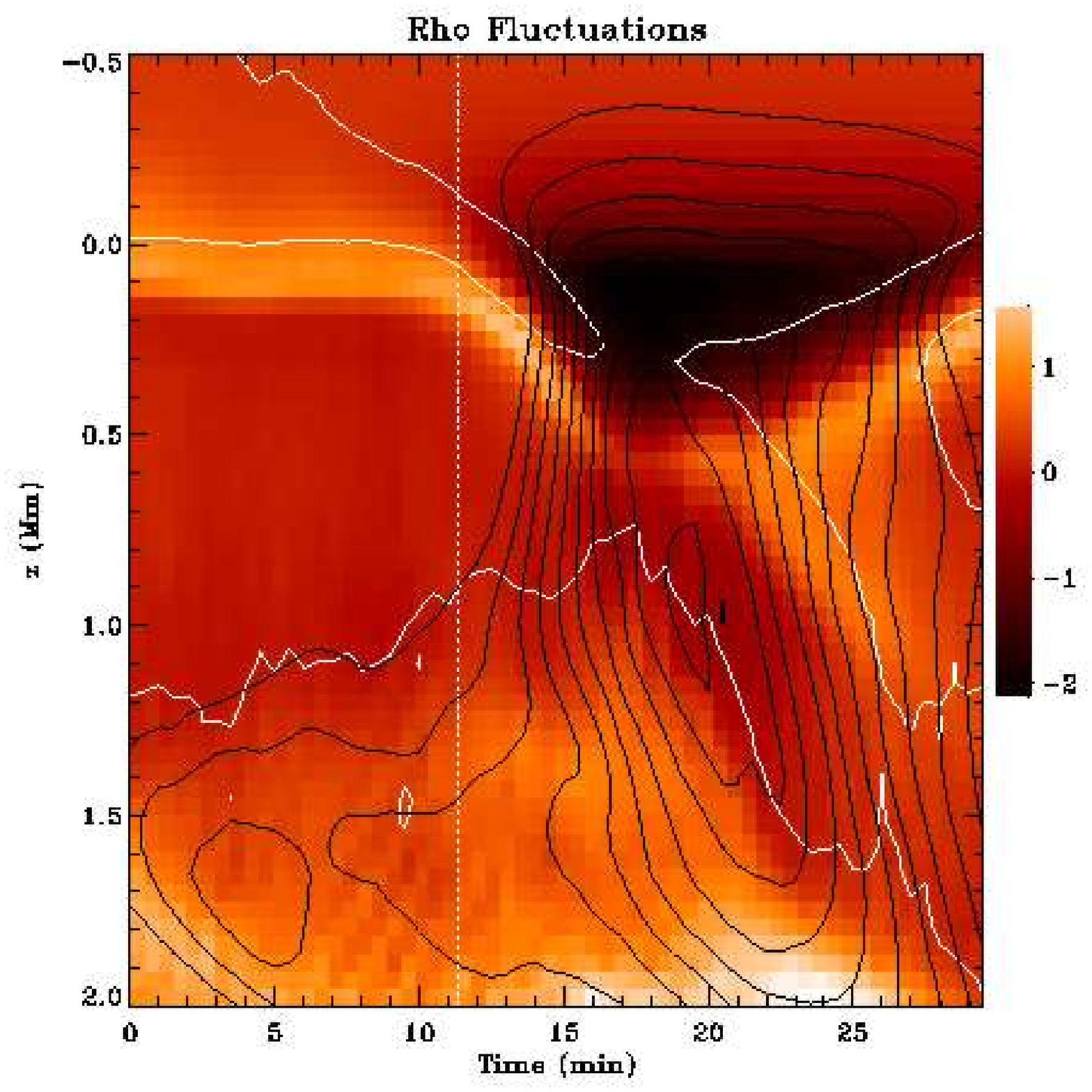}
  }
  \vspace*{0.2cm}
  \hbox{
  \epsfxsize=6cm
  \epsffile{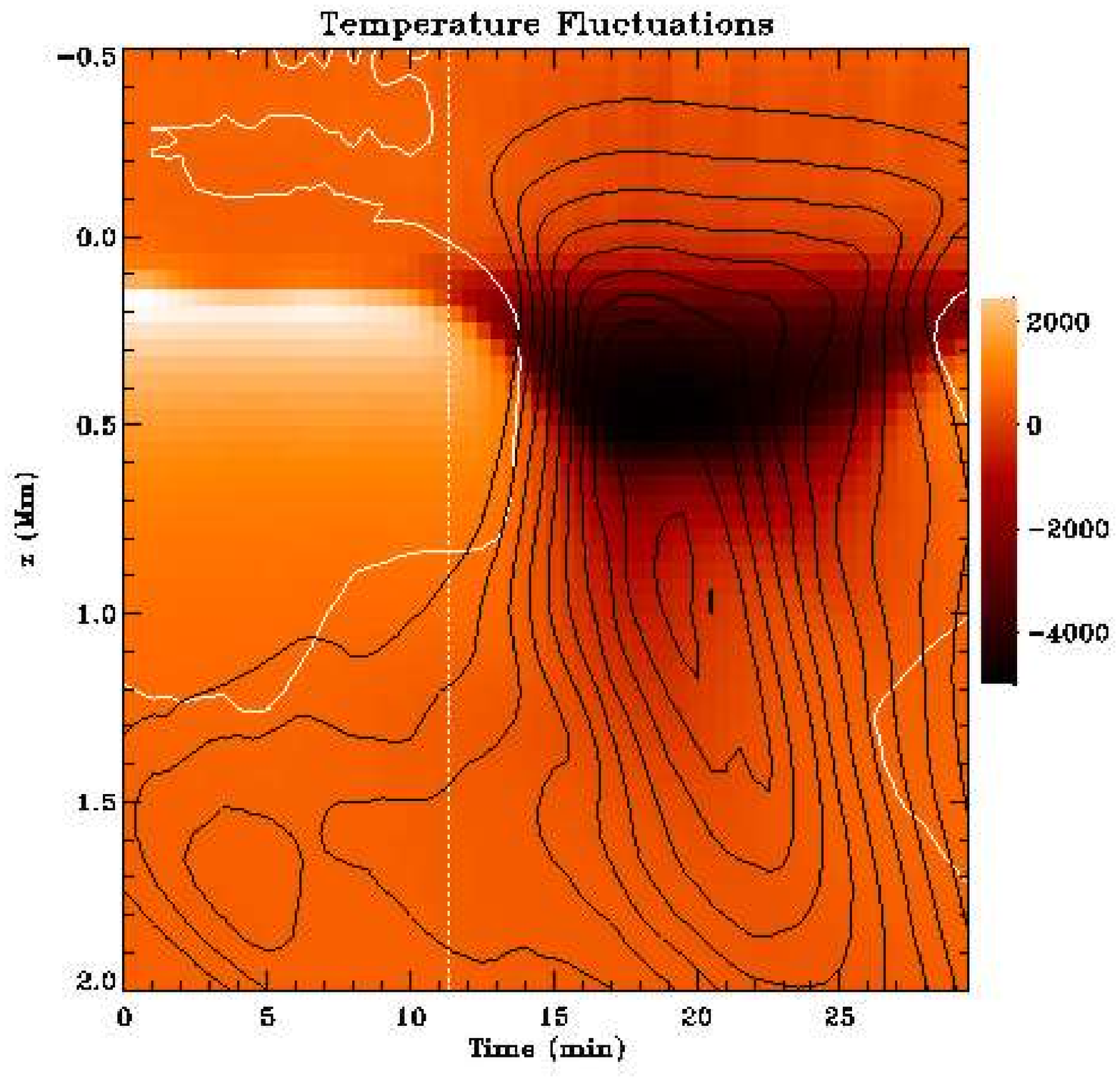}
  \hspace*{0.3cm}
  \vspace*{-0.2cm}
  \epsfxsize=6cm
  \epsffile{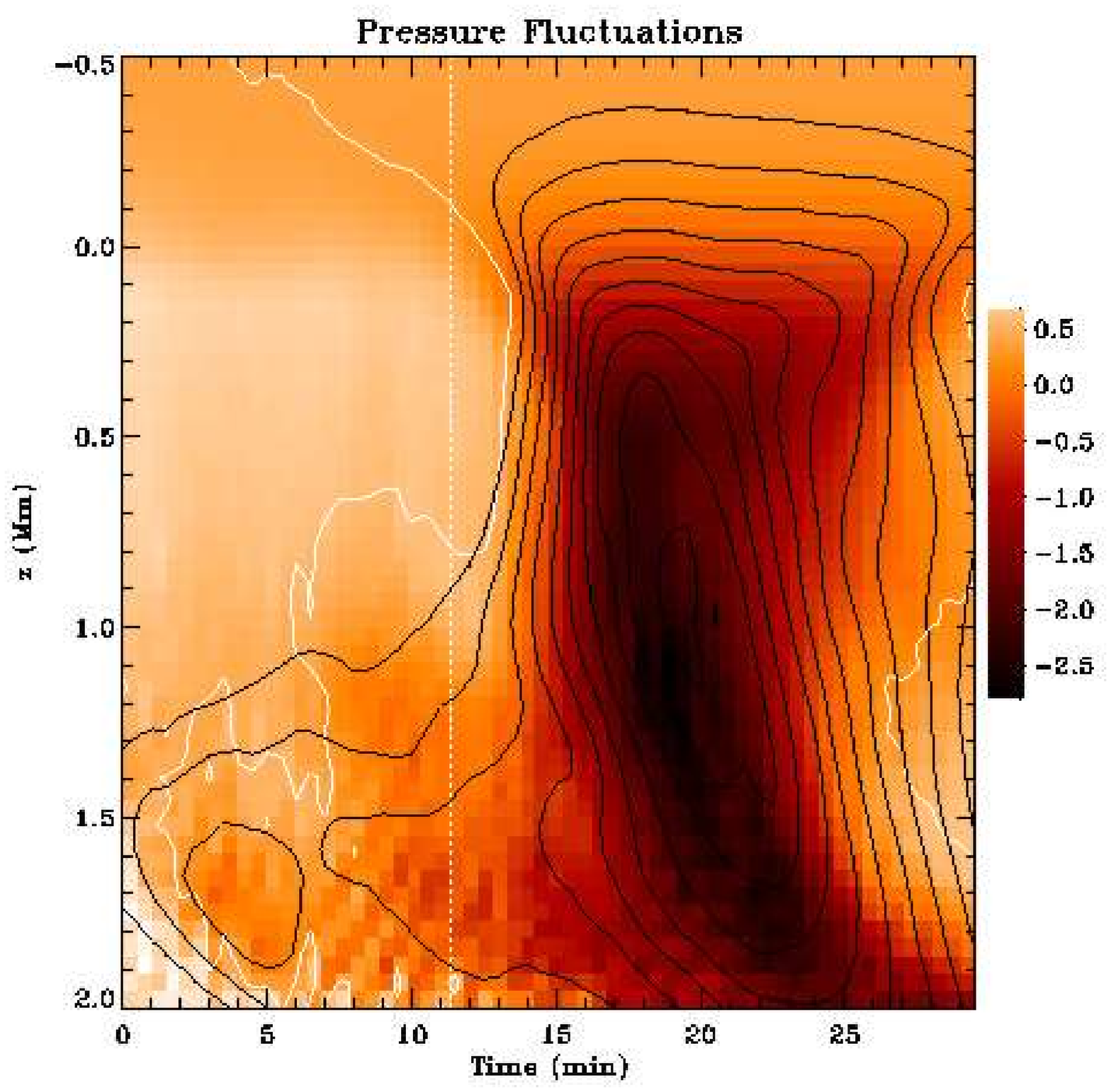}
  }
  \caption[]{\label{Vz-t-z}{\small
  Evolution of the vertical velocity (top left),
  density fluctuations (top right),
  temperature fluctuations (bottom left),
  and pressure fluctuations (bottom right)
  on a vertical line through the center of the
  micropore.  White regions are the
  downward velocities and dark regions are the upward velocities.  White
  lines indicate zero vertical velocity.  Dark contours are the magnetic
  field strength.}
  }
\end{figure}

The strong field concentrations persist for hours in the
same region, much longer than the micropores, which form and disappear.

We can also gain insight into micropore evolution by looking at how
variables on a line through the center of the micropore evolve
in time.  Fig.~\ref{Vz-t-z} shows change with time of the vertical
velocity on a line through the center of the micropore.   Initially
there is an upflow in the granule at the location of the future micropore.  As the
field concentrates this switches to a large downward velocity,
which starts at the surface and propagates downward.  Again as the
magnetic field disperses there is a large downward velocity.  The
downflow as the micropore forms has higher than average density and
evacuates the flux tube, which produces a low density region
near the surface in the micropore (Fig.~\ref{Vz-t-z}).  As the micropore
disappears, higher density material moves back toward the surface,
bringing the density there back to average.  Originally the granule is hot.
As the micropore forms the region cools off (Fig.~\ref{Vz-t-z}) because
not as much energy is being transported to the surface and while cooling
from the surface continues unabated.  As the field disperses the fluid
gets heated back up again and the micropore vanishes.  Before the
micropore forms the pressure in the granule is higher than the
surroundings pushing the diverging flow
(Fig.~\ref{Vz-t-z}).  As the magnetic field gets concentrated,
of course, the gas pressure
goes down so as to maintain approximate pressure balance in horizontal
directions with its surroundings.  As the field disperses the pressure
comes back up again and produces a large downward pressure gradient in the
micropore.



\acknowledgments
This work was supported in part by NASA grant NAG
5-9563, NSF grant AST 9819799, and the Danish Research
Foundation, through its establishment of the Theoretical Astrophysics
Center.  The calculations were performed at the National Center for
Supercomputer Applications, which is supported by the National Science
Foundation, at Michigan State University and at UNI$\bullet$C,
Denmark. 

\end{document}